\documentclass[a4paper,11pt]{article}
\pdfoutput=1 

\usepackage{jcappub} 

\usepackage[T1]{fontenc} 
\usepackage{mathrsfs,amsmath}
\usepackage[noend]{algpseudocode}
\usepackage{algorithm}
\usepackage{scalerel}
\usepackage{tensor}
\usepackage{color, colortbl}
\definecolor{light-gray}{gray}{0.95}
\definecolor{darkred}{rgb}{0.7,0,0}

\usepackage[export]{adjustbox}

\title{Small Field models with ACTPol and BICEP3 data - Likelihood analysis.}

\author[\bold{a},1]{Ira Wolfson,\note{Corresponding author.}}
\author[\bold{b}]{Utkarsh Kumar,}
\author[\bold{b,c}]{Ido Ben-Dayan,}
\author[\bold{d}]{and Ram Brustein}

\affiliation[\bold{a}]{International School for Advanced Studies (SISSA), Data Science Excellence Department, Via Bonomea 265, 34136 Trieste, Italy
}

\affiliation[\bold{b}]{Physics Department, Ariel University, Ariel 40700, Israel
}

\affiliation[\bold{c}]{Department of Physics, University of California, Berkeley, CA 94720, USA
}
\affiliation[\bold{d}]{Department of Physics, Ben-Gurion University, Beer-Sheva 84105, Israel
}

\emailAdd{iwolfson@sissa.it}
\emailAdd{ido.bendayan@gmail.com}
\emailAdd{}

\abstract{ We perform a Bayesian analysis for small field models of inflation, using the most recent datasets produced by Planck`18, ACTPol, and BICEP3. We employ Artificial Neural Networks (ANN) to perform analyses with model coefficients, instead of their proxy slow-roll parameters. The ANN connects the models with their projected scalar index $n_s$ and index running $\alpha$, in lieu of the less accurate Lyth-Riotto expressions. We recover the most likely coefficients for a sixth degree polynomial inflationary potential, which yields a tensor-to-scalar ratio $r\lesssim 0.03$. We do so for the case of joint Planck and ACTPol datasets, and for each dataset alone. The BICEP3 data is included in all three analyses. We show that these models are likely, with coefficients that are tuned to about $\Delta\gtrsim 1/60$. Curiously, we also find a significant tension between ACTPol and Planck datasets, which we try to account for. }

\begin{document}
\maketitle
\flushbottom
\section{Introduction}
The observation that non-causal regions of the Cosmic Microwave Background (CMB) are in thermal equilibrium \cite{Penzias:1965wn}, have given rise to several competing theories for the cosmic origin. Chief among them is the theory of inflation \cite{Starobinsky:1980te,Sato:1980yn,Guth:1980zm,Linde:1981mu,Albrecht:1982wi}, an epoch of accelerated expansion of the Universe. In addition to solving the standard hot Big Bang model problems, Inflation predicts inhomogeneities and anisotropies in the CMB with a nearly flat spectrum \cite{Mukhanov:1981xt,Hawking:1982cz,Starobinsky:1982ee,Guth:1982ec,Bardeen:1983qw}.
It also predicts non-Gaussianities which vanish to first order \cite{Bartolo:2004if}, and the production of a Gravitational Wave (GW) signal \cite{Grishchuk:1974ny,Starobinsky:1979ty,Abbott:1984fp}. Inflation is well supported by our current cosmological observations \cite{Bennett:1996ce,Hinshaw:2012aka,Akrami:2018odb}.\\

In the inflationary slow-roll paradigm, a quasi-constant energy density powers the accelerated expansion. However, the nature of the mechanism still eludes us. Thus a plethora of different models are suggested and examined (c.f. \cite{Martin:2013tda,Martin:2014lra} ), of which the simplest is implemented with a single inflaton $\phi$ field slowly-rolling down the inflationary potential $V(\phi)$.\\

 The Lyth bound \cite{Lyth:1996im} connects the GW signal amplitude with the slope of the potential. Thus, for the last few decades the data seem to have favored large-field models, in which the field excursion is more than a few Planck units $\Delta\phi> 2\sim3 \;\left(M_{\mathrm{Pl}}\right)^{-1/2}$. But recent developments, theoretical \cite{Garg:2018reu,Ben-Dayan:2018mhe, Palti:2019pca,Kehagias:2019iem}, phenomenological \cite{Ben-Dayan:2009fyj,Hotchkiss:2011gz} and computational \cite{Wolfson:2019rwd} have shown the small-field models with $\Delta\phi \simeq 1$ are at least as likely as large-field models. \\

Meanwhile, the available observational data has multiplied many folds \cite{Bennett:1996ce,Hinshaw:2012aka,Akrami:2018odb,ACT:2020gnv}. So we find ourselves with an embarrassment of riches. We have more data than we can hope to analyze directly. We therefore make use of increasingly complex computational tools to mine insights and test our models.\\

\begin{figure}[!th]
    \centering
    \includegraphics[width=0.8\textwidth]{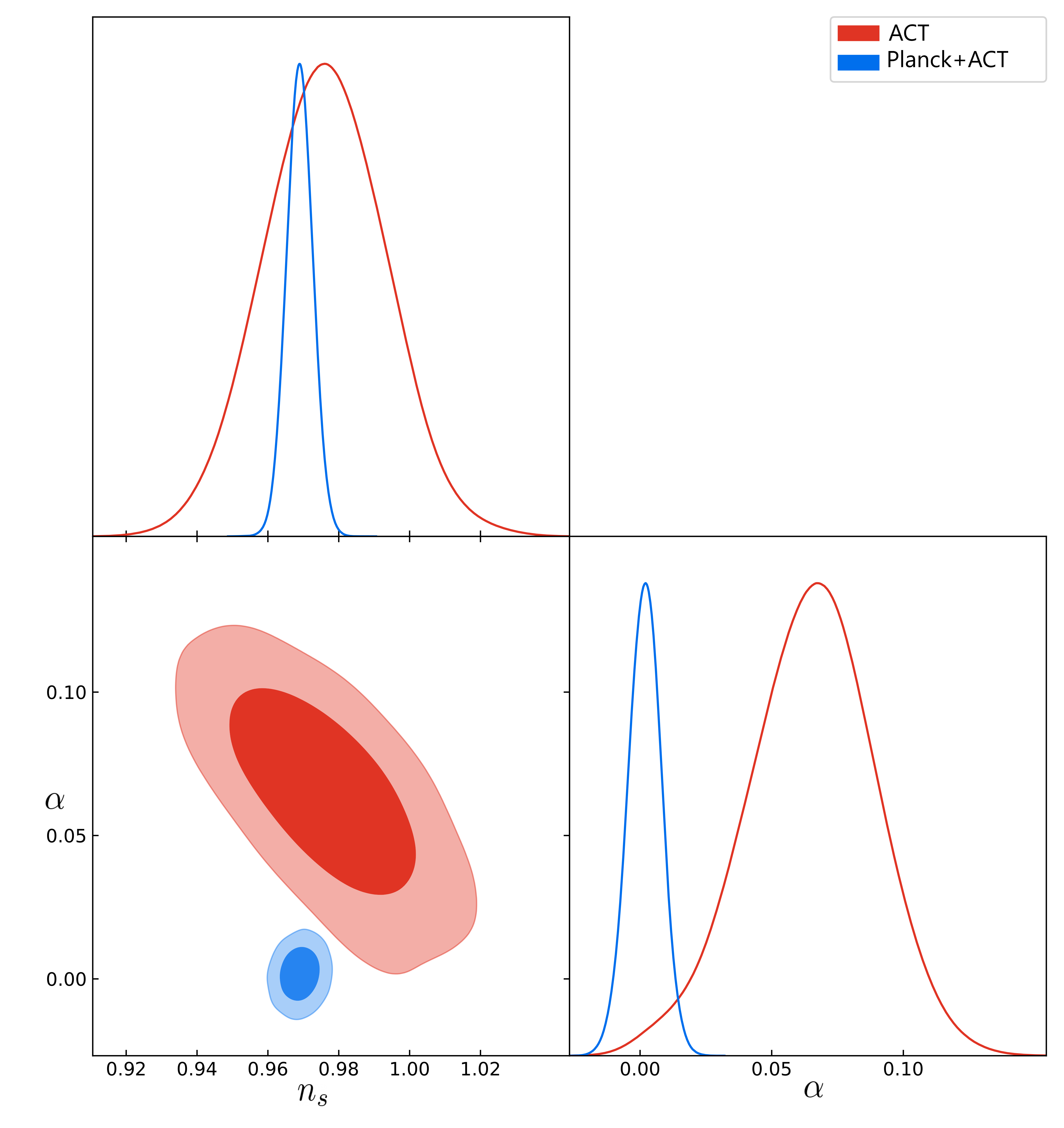}
    \caption{An MCMC analysis of ACTPol data, both combined with Planck data (blue curves), and stand-alone (red curves). There is a marked difference in both scalar index $n_s$, and index running $\alpha$. The value of $\alpha$ excludes zero with $\sim 2.8 \sigma$.}
    \label{fig:ACTpolMCMC}
\end{figure}

Recent ACTPol results \cite{ACT:2020gnv} suggest a spectral tilt $n_s>1$ which is several standard deviations away from the inferred Planck value of $n_s\simeq 0.965$. Since ACTPol is sensitive mostly to higher multipoles, and since Planck's data is much richer, an immediate "fix" is adding a running of the spectral index,  denoted by $\alpha$. However, even including $\alpha$ the ACTPol data is still $\sim 2.8\sigma$ away from the Planck-ACTPol analysis, see figure~\ref{fig:ACTpolMCMC}. Hence, a non-trivial scale dependence is called for. A natural candidate is a polynomial small field model with enhanced GW signal, where enhanced scale dependence is an inherent feature \cite{Ben-Dayan:2009fyj,Wolfson:2019rwd,Wolfson:2016vyx,Wolfson:2018lel}.   
We use the recently published ACTPol \cite{ACT:2020gnv} and BICEP3 \cite{BICEP:2021xfz} data to extract the likelihood of small-field models of a polynomial type. The specific models we use all yield a tensor-to-scalar ratio $r\simeq 0.03$. Given the latest observational results constraining $r\lesssim 0.032$ \cite{Tristram:2021tvh}, these potentials can be tested in the very near future. \\
~Probing the smaller scales of the primordial power spectrum, i.e. more e-folds of inflation has virtue of its own. It may also include features or deviations not necessarily parameterized by $\alpha$. Several probes which are not based on CMB temperature or polarization anisotropies come to mind, Galaxy Surveys such as MegaMapper \cite{Sailer:2021yzm}, spectral distortions of the CMB black body spectrum \cite{Chluba:2012we}, weak lensing of supernovae \cite{Ben-Dayan:2013eza,Ben-Dayan:2014iya,Ben-Dayan:2015zha}, the number of relativistic degrees of freedom $N_{eff}$ and even gravitational wave interferometers \cite{Ben-Dayan:2019gll}. Combining data from such probes will place further constraints on $\alpha$. We did not incorporate these additional probes into our analysis, but we shall come back to them in the discussion section.

The paper is organized as follows. First we discuss the Bayesian analysis of the ACTPol data when combined with the Planck'18 and BICEP3 databases, versus when it is analysed on its own. We show that, when including the running of the scalar index in the analysis, there is a marked tension in the resulting primordial power spectra. We proceed to introduce our models, and the computational tools used for this analysis. Finally we present and discuss the results and their possible implications.\\

~We work with units in which $\hbar=c=1$, the reduced Planck mass is set to $1$ and the metric signature is $(-,+,+,+)$.
\begin{figure}
    \centering
    \includegraphics[width=0.9\textwidth]{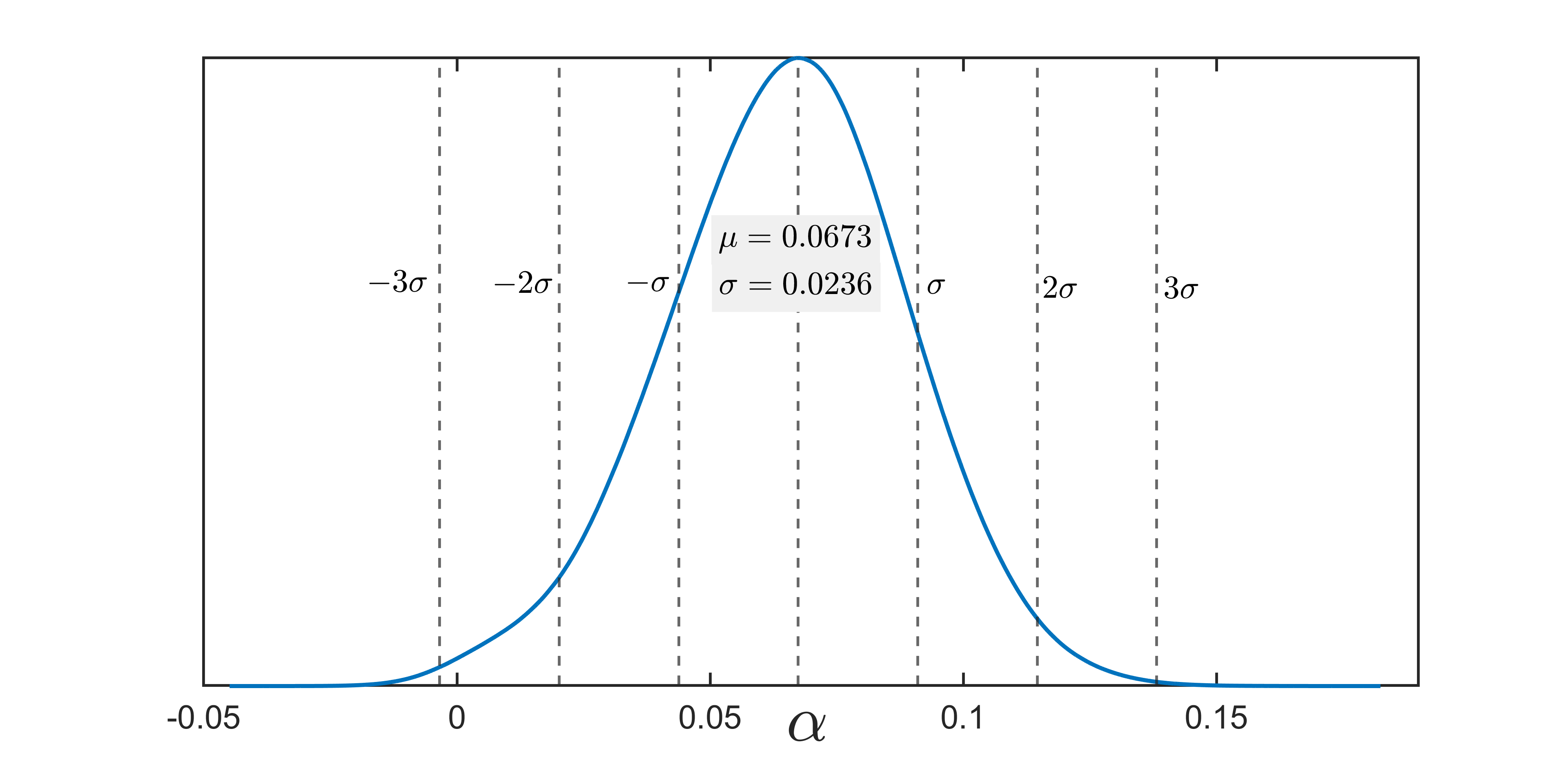}
    \caption{A 1D graph of the index running $\alpha$ after marginalization over the other parameters. The mean value and $\sigma$-values are plotted vertically. $\alpha$ seems to be positive, and removed from 0 at a level of $\sim 2.8\;\sigma$.}
    \label{fig:alpha_Sigma}
\end{figure}
\section{MCMC analysis of ACTPol data}
Connecting the CMB power spectrum to the spectrum of quantum perturbations during inflation is an inverse problem. These quantum perturbations are represented in the form of the primordial power spectrum (PPS):
\begin{align}
    P(k)=A_s\left(\frac{k}{k_0}\right)^{n_s-1 +\tfrac{\alpha}{2}\ln\left(\tfrac{k}{k_0}\right)+...},
\end{align}
in which $k$ is the wave-number, $n_s$ is the scalar index, $\alpha$ is its running, and $A_s$ is the amplitude of the power spectrum at the pivot scale $k_0$.
These perturbations then go through physical evolution to produce the CMB power spectrum.
We use the cosmological Markov-Chain-Monte-Carlo (MCMC) code cobaya \cite{Torrado:2020dgo} with the CAMB \cite{Lewis:1999bs} component for line-of-sight analysis to recover the most likely primordial power spectrum given the observational data. \\
~We performed several different analyses, using the ACTPol DR4 data set which is publicly available \cite{ACT:2020gnv,ACT:2020frw}, both in combination with the Planck data and with no additional data sets. In these early analyses we did not add the BICEP3 dataset. We probed different PPS degrees of freedom $(n_s,\alpha)$ etc. As was shown in \cite{ACT:2020gnv}, the scalar index $n_s$ is markedly different between ACTPol+Planck analysis and ACTPol stand-alone analysis. This difference may be due to physical effects, as the ACTPol CMB data is calibrated to agree with Planck in the $600<l<1800$ region. Planck analysis implies a scalar index of $\sim 0.965$ in the low $l$'s, where Planck constrains the angular power spectrum in the $300<l<1800$ range. However, the stand-alone analysis of the ACTPol data recovers an $n_s\gtrsim 1$, in the smaller scales of $l\gtrsim 1800$, where ACTPol constrains the power spectrum better (\cite{ACT:2020gnv},Fig.~23). If one believes both analyses, one must conclude that the spectral index running is significant. Indeed, the analysis in \cite{ACT:2020gnv} shows the different results of MCMC analyses with the running of the spectral index $\alpha$ as a free parameter. When combining ACTPol with either WMAP or Planck data, $\alpha$ is always within $\sim 1\sigma$ of $\alpha=0$. However, using ACTPol data alone, we recover a significant positive $\alpha$ that is removed from $\alpha=0$ by $\sim 2.8 \sigma$ (see figures~\ref{fig:ACTpolMCMC},\ref{fig:alpha_Sigma}). We posit, that by combining the ACTPol data with Planck or WMAP, we effectively dilute the ACTPol signal in the much richer data of Planck and WMAP. This explains the result of the combination being more similar to Planck/WMAP when combined with them. This may also explain the disparity in uncertainties between ACTPol alone, and when combined with the earlier CMB satellite experiments.
\section{Setup}
The action for single scalar field inflation is given by:
\begin{align}
    \mathcal{A}=\int d^4x \sqrt{-g} \left[-\frac{\mathcal{R}}{2}-\frac{1}{2}\partial_\mu \phi \partial^\mu \phi -V(\phi)\right],
\end{align}
in which $g$ is the trace of the metric, $\mathcal{R}$ is the Ricci scalar and $\phi$ is the inflaton. In this work we study small-field potentials of the form
\begin{align}
    V(\phi)=1 +\sum_{n=1}^{6}a_n \phi^n,
\end{align}
embedded in an FRW metric:
\begin{align}
    g_{\mu\nu}=-\delta_{0\mu}\delta_{0\nu}+a^2(t)\delta_{ij},
\end{align}
where $a(t)$ is the time-dependant scale factor.\\
\begin{figure}[!h]
    \centering
    \includegraphics[width=0.8\textwidth]{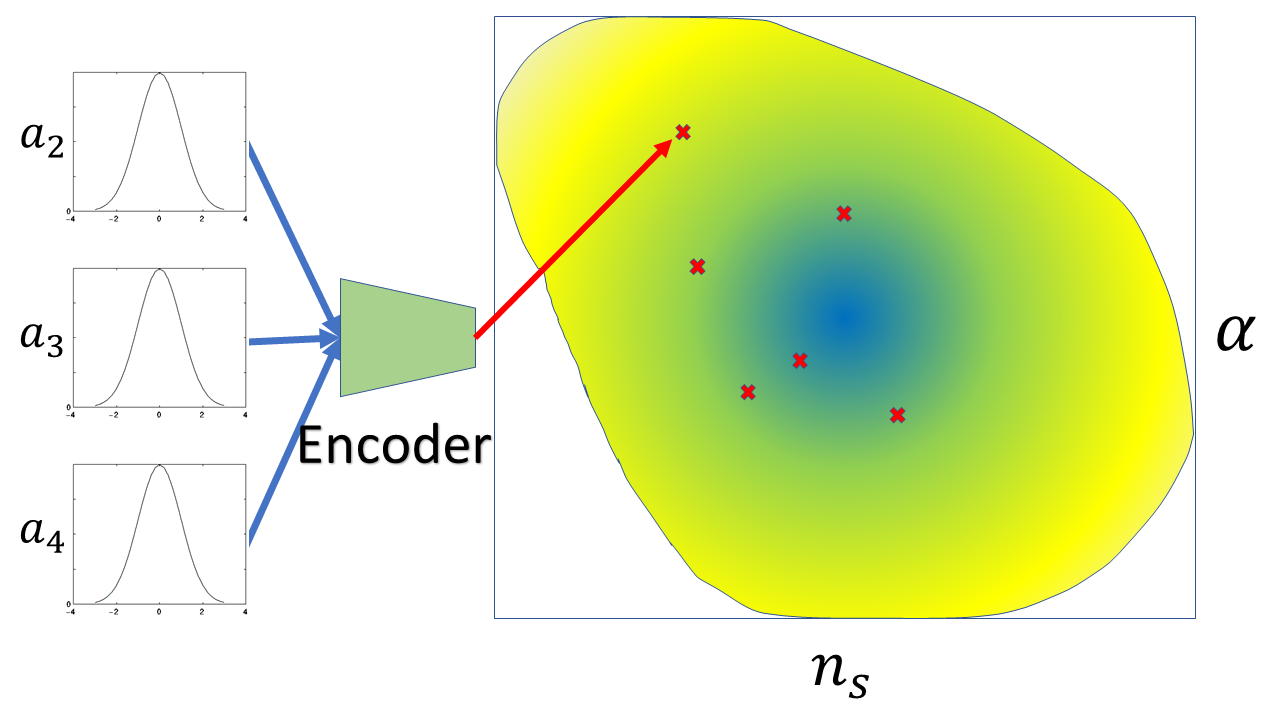}
    \caption{Using the encoder we generate samples from the coefficient phase-space priors, triangulate them in the observable phase-space and generate Markov chains. We thus recover the posterior of the coefficients phase-space.}
    \label{fig:MCMC+Encoder}
\end{figure}

~This type of models can be viewed as a perturbed variant of an extended hilltop model of the type:
\begin{align}
    V(\phi)=1-a_k\phi^k,
\end{align}
for a specific $k\geq 2$. We use models of $k=6$ which we constrain by the small-field requirement $\Delta\phi=1$. The first coefficient $a_1$ is given by:
\begin{align}
    a_1= -\sqrt{\frac{r}{8}},
\end{align}
where $r$ is the tensor-to-scalar ratio. Thus, requiring a specific GW amplitude in the CMB further fixes the model. Finally, a `soft' requirement of $N=60$ leaves only three degrees of freedom for a model of this class. \\
Thus to write the model explicitly we have:
\begin{align}
    V(\phi)=1 -\sqrt{\frac{r}{8}}\phi +a_2\phi^2 +a_3\phi^3 +a_4\phi^4 +a_5{}_{\left(a_{1-4},a_6\right)}\phi^5 +a_6{}_{\left(a_{1-4}\right)}\phi^6,
\end{align}
where the coefficients $\{a_2,a_3,a_4\}$ are free parameters and $\{a_5,a_6\}$ are constrained as explained above, and are thus functions of the other coefficients.\\
These models have already been studied in \cite{Wolfson:2019rwd}, and have been shown to be promising candidates for inflaton models. These models also adhere to the Lyth bound \cite{Lyth:1996im}, while still producing a high GW signal. Specifically, in this work we set $r=0.03$, and derive $a_1$ accordingly.\\
\begin{figure}[!h]
    \centering
    \includegraphics[width=0.9\textwidth]{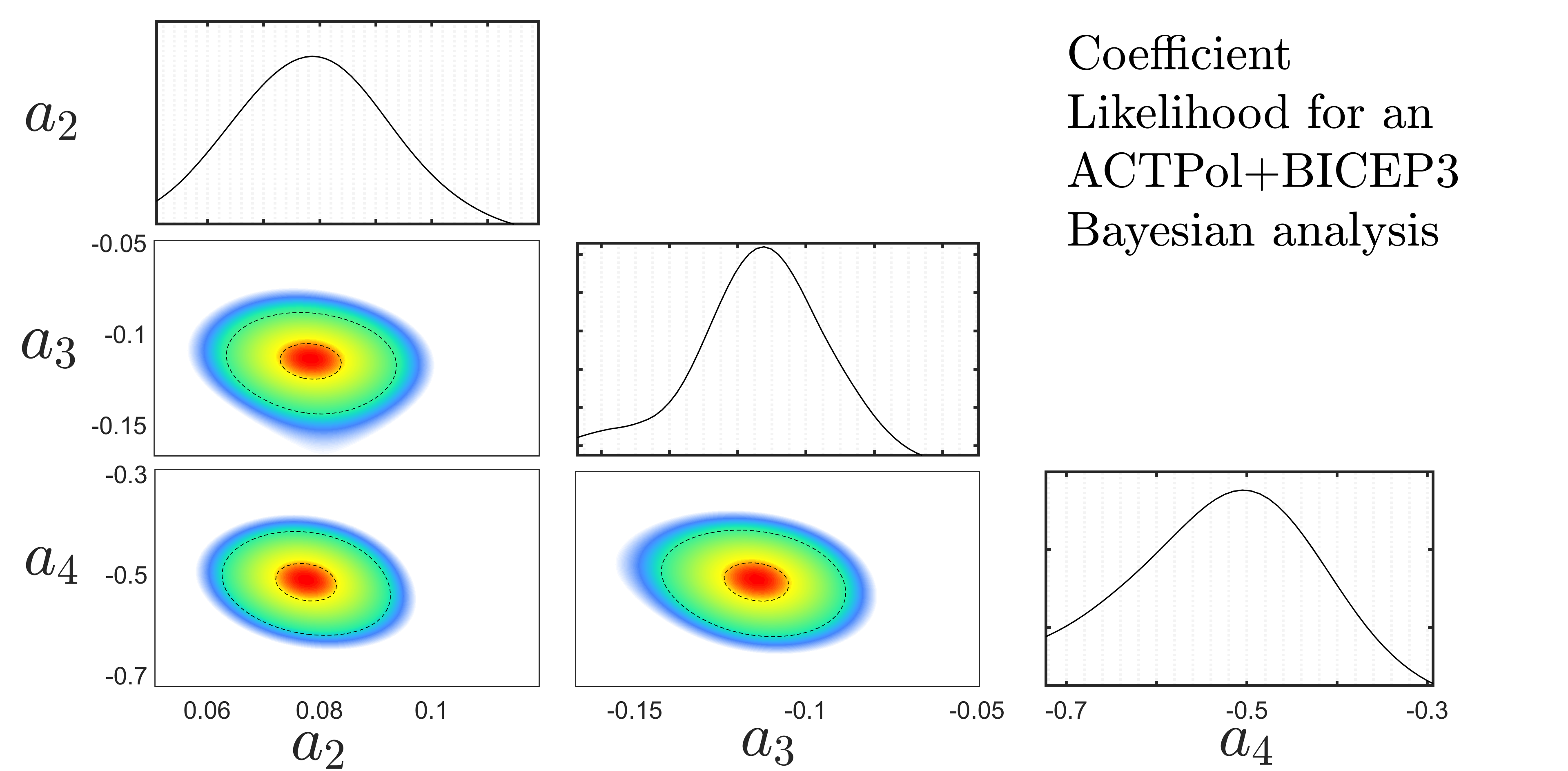}
    \caption{An MCMC analysis of coefficient trio $\{a_2,a_3,a_4\}$. This analysis was performed with an ANN that predicts $n_s$ and $\alpha$. The underlying joint distribution function is a result of a Bayesian analysis using ACTPol and BICEP datasets.}
    \label{fig:ACT_MCMC}
\end{figure}
\section{Methods}
\subsection{AutoEncoders}
The use of encoders is not new. Encoders are used in many aspects of communication and signal processing. One prevalent use is noise reduction and faithful signal reconstruction. Another use is data compression, where a varied input in the original space may be encoded onto very few underlying variables, i.e the so-called `latent space' \cite{Santurkar:2018,Chen:2016}. In recent years Neural Networks (NNs) are increasingly employed as efficient non-linear encoders (for instance \cite{Khemakhem:2019}). These are usually called AutoEncoders (AE), or Variational AutoEncoders (VAE). The general idea is training a NN to faithfully encode and then decode data, such that the loss of information between input and reconstructed output ideally vanishes, or is at least minimal. A concise review of AEs and VAEs can be found in \cite{kingma:2019}.\\

~ In \cite{Wolfson:2021zsw} it was shown that the PPS produced by small field models cannot be analytically approximated to less than $1\%$ error consistently. Thus in the absence of a simple function that accurately predicts the cosmological observables $n_s$ and $\alpha$, we make use of an encoder. Given the numerical values of potential coefficient, this encoder is trained to triangulate the correct scalar index and index running the inflationary evolution of this inflationary potential yields. The general idea is shown in figure~\ref{fig:MCMC+Encoder}. \\

~ Specifically, After extraction of the likelihood functions using cobaya, we use an MCMC engine to retrieve the posterior distributions of the coefficients $\{a_2,a_3,a_4\}$ for models with $r=0.03$. We then filter out models that yield $n_s$ and $\alpha$ that significantly diverge from the physical values observed. These data are quintuples of the form $(a_2,a_3,a_4,n_s,\alpha)$. Instead of a classic AE, we use only the encoder part, since we treat the data as the ground truth. \\

~ We toggle over many architectures of encoding networks, in search of the minimal architecture NN, that yields an acceptable statistical error of less than $0.01\%$. The process and the exact architecture used are detailed in \cite{Wolfson:2021zsw}.

\subsection{MCMC of potential parameters made possible}
\begin{figure}[!h]
    \centering
    \includegraphics[width=0.9\textwidth]{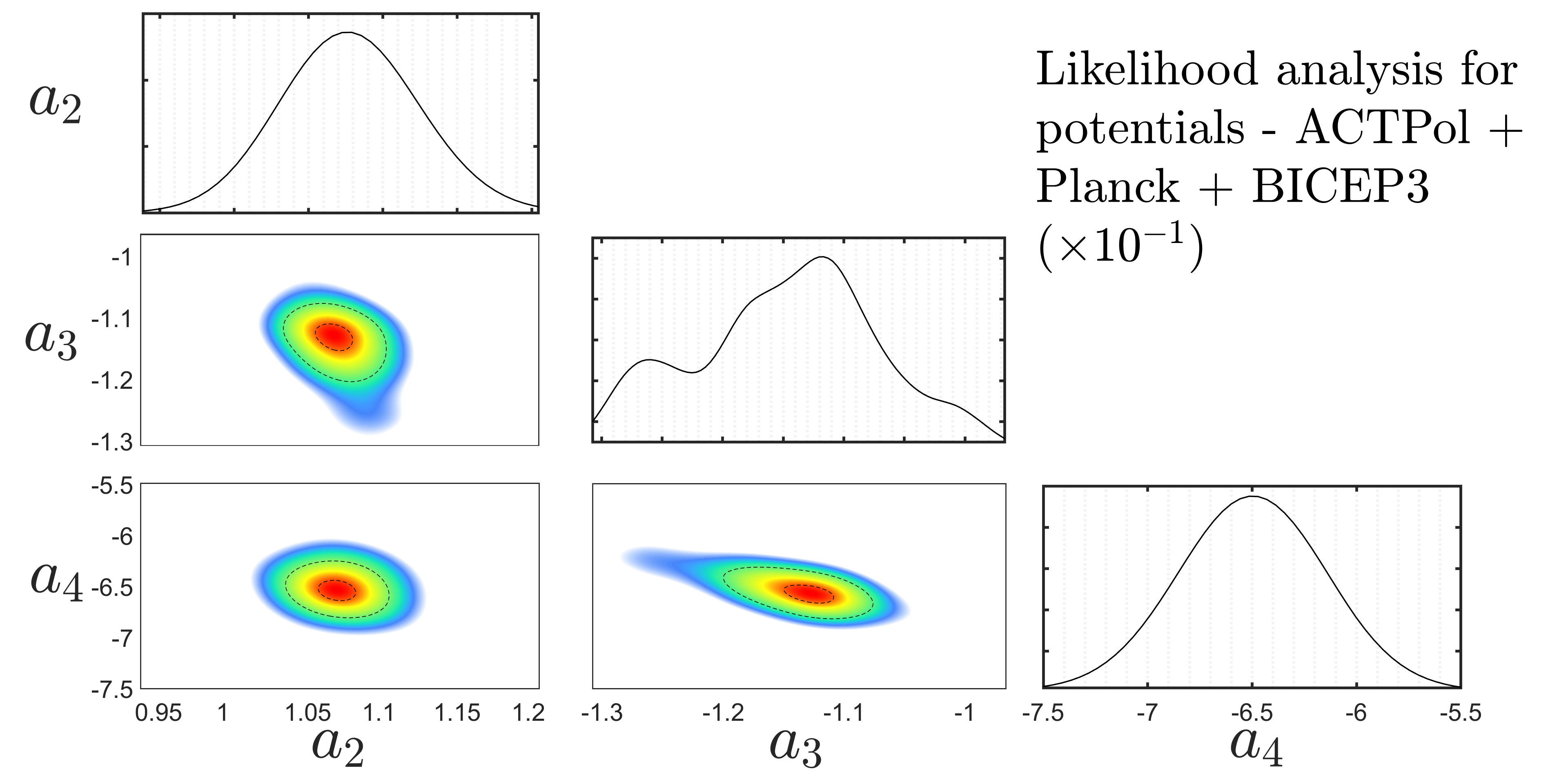}
    \caption{An MCMC analysis of coefficient trio $\{a_2,a_3,a_4\}$. This analysis was performed with an ANN that predicts $n_s$ and $\alpha$. The underlying joint distribution function is a product of a Bayesian analysis with the ACTPol, Planck, and BICEP datasets. While the likelihood of $a_3$ looks to be a mixture of 3 Gaussians, there is still a clear peak around $a_3\sim -0.0115$.}
    \label{fig:ACT+PlanckMCMC}
\end{figure}
 Introducing a full numerical calculation in an MCMC engine for each sampled set of coefficients was unfeasible in terms of computational run-time. Even a $2$-second calculation for one potential, amounts to an addition of $\sim 2$ months for every reasonable MCMC run. Thus, we were previously reduced to manufacturing a large sample of inflationary potentials, and `lifting' the underlying CMB observable likelihoods by superimposing the calculated PPS onto the 2D observable posteriors. While that method may yield correct results, it relies on simplifying assumptions such as a vanishingly small paired covariance of the observables. A detailed explanation of our prior method is found in \cite{Wolfson:2018lel}.\\

~ However, the use of a NN encoder allows us to calculate observables from potentials in computationally feasible times.  While this computation is not reliable with every single potential as the maximal error for these predictions is still large $\left(\sim0.5\%\right)$, the statistical mean is of order $0.01\%$. Thus a statistical MCMC analysis is made possible.\\

~When sampling a trio $(a_2,a_3,a_4)$, we do so from a uniform distribution spanning the minimum to maximum values for each coefficient, of the data set which the NN was trained with. The trio is fed to the NN, which predicts the observable it yields. These observables are then triangulated in the 2D scalar index - index running $(n_s,\alpha)$ phase-space, and the likelihood of the specific observables is used in the accept/reject component of the MCMC sampler. Convergence is measured by employing the Gelman–Rubin diagnostic \cite{Gelman:1992zz} across all simulated chains. The complete procedure, and the details of the NN are documented in \cite{Wolfson:2021zsw}. But for completeness we outline it here:\\
\begin{algorithm}[!h]
\caption{Sampling over Potential Coefficients (with NN):}
\begin{algorithmic}[1]
\Require{Current point $x_i$ and it's associated likelihood $\mathscr{L}^i$.}

    \State { $\boldsymbol{x'} \gets \left(a_2,a_3,a_4\right)$} \Comment{ Sample $a_i\sim U\left(min(a_i),max(a_i)\right)$  }
    \State $\boldsymbol{\left(n'_s,\alpha'\right)} \gets \mathscr{F}_{NN}(x')$ \Comment{$\mathscr{F}_{NN}:\mathscr{R}^3 \rightarrow \mathscr{R}^2$}
    \Statex \Comment{(see figure~\ref{fig:MCMC+Encoder}).}
    \State $\boldsymbol{\mathscr{L}'} \gets J(n'_s,\alpha')$ \Comment{$\mathscr{L}'$ gets the joint probability of $n'_s$ and $\alpha'$ }
    \State $Z \gets \tfrac{\mathscr{L}'_{x'}}{\mathscr{L}^i_{x_i}}$ \Comment{Evidence ratio}
    \State $M \gets \tfrac{q(x'|x_i)}{q((x_i|x')}$
    \State $u \gets U[0,1]$\Comment{u is uniformly distributed on [0,1]}
    \If{$u\leq \tfrac{Z}{M}$}
        \State $\boldsymbol{x_{i+1}}\gets \boldsymbol{x'}$\Comment{Accept}
    \Else
        \State $\boldsymbol{x_{i+1}}\gets \boldsymbol{x_i}$\Comment{Reject}
    \EndIf
\end{algorithmic}
\end{algorithm}
\newpage
In this way we produce Markov chains, which we then marginalize over to find the posterior distributions of the different coefficients $a_i$.

\section{Results}
\begin{figure}[!h]
    \centering
    \includegraphics[width=0.9
    \textwidth]{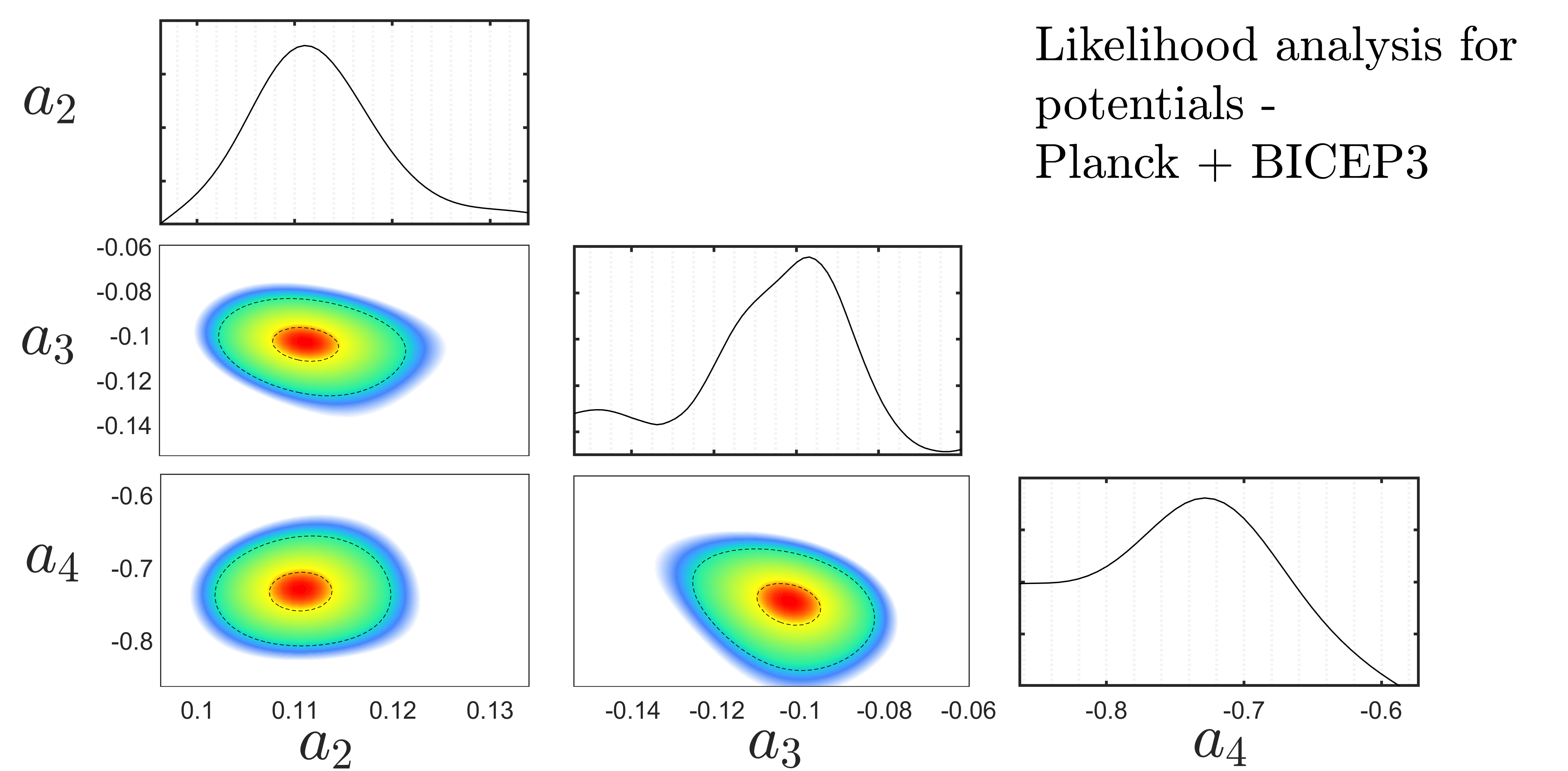}
    \caption{An MCMC analysis of coefficient trio $\{a_2,a_3,a_4\}$. This analysis was performed with an ANN that predicts $n_s$ and $\alpha$. The underlying joint distribution function is a result of a Bayesian analysis using Planck and BICEP datasets. Due to tighter constraint on the target values, the model's non-linearities become more pronounced. Hence the skewness. The lines correspond to the $68\%$ and $95\%$ probability density lines.}
    \label{fig:PLANCK-BICEP}
\end{figure}
We see that both for the ACTpol, Planck and BICEP joint data presented in figure~\ref{fig:ACT+PlanckMCMC}, as well as for ACTpol data joint only with BICEP, shown in figure~\ref{fig:ACT_MCMC}, there are sets of coefficients that give rise to the desired primordial power spectra. For the case of ACTPol,Planck and BICEP joint data the primary modes of the marginal likelihoods on the coefficients are given by:
\begin{align}
    a_2=0.108\pm 0.011\hspace{10pt} a_3=-0.114\pm 0.014\hspace{10pt}a_4=-0.650 \pm 0.085\;.
\end{align}
However, for ACTPol and BICEP without Planck, the results are given by:
\begin{align}
    a_2=0.078\pm 0.034\hspace{10pt} a_3=-0.112\pm 0.047\hspace{10pt}a_4=-0.496\pm 0.219\;.
\end{align}
Finally, for completeness we ran the same analysis with the Planck'18 database combined with the BICEP3 database, but without the ACTPol data. This analysis is shown in figure~\ref{fig:PLANCK-BICEP}, and yields the following most likely coefficients:
\begin{align}
    a_2=0.111\pm 0.016\hspace{10pt} a_3=-0.099 \pm 0.033\hspace{10pt}a_4=-0.723 \pm 0.177\;.
\end{align}

\begin{figure}[!ht]
    \centering
    \includegraphics[width=0.9\textwidth]{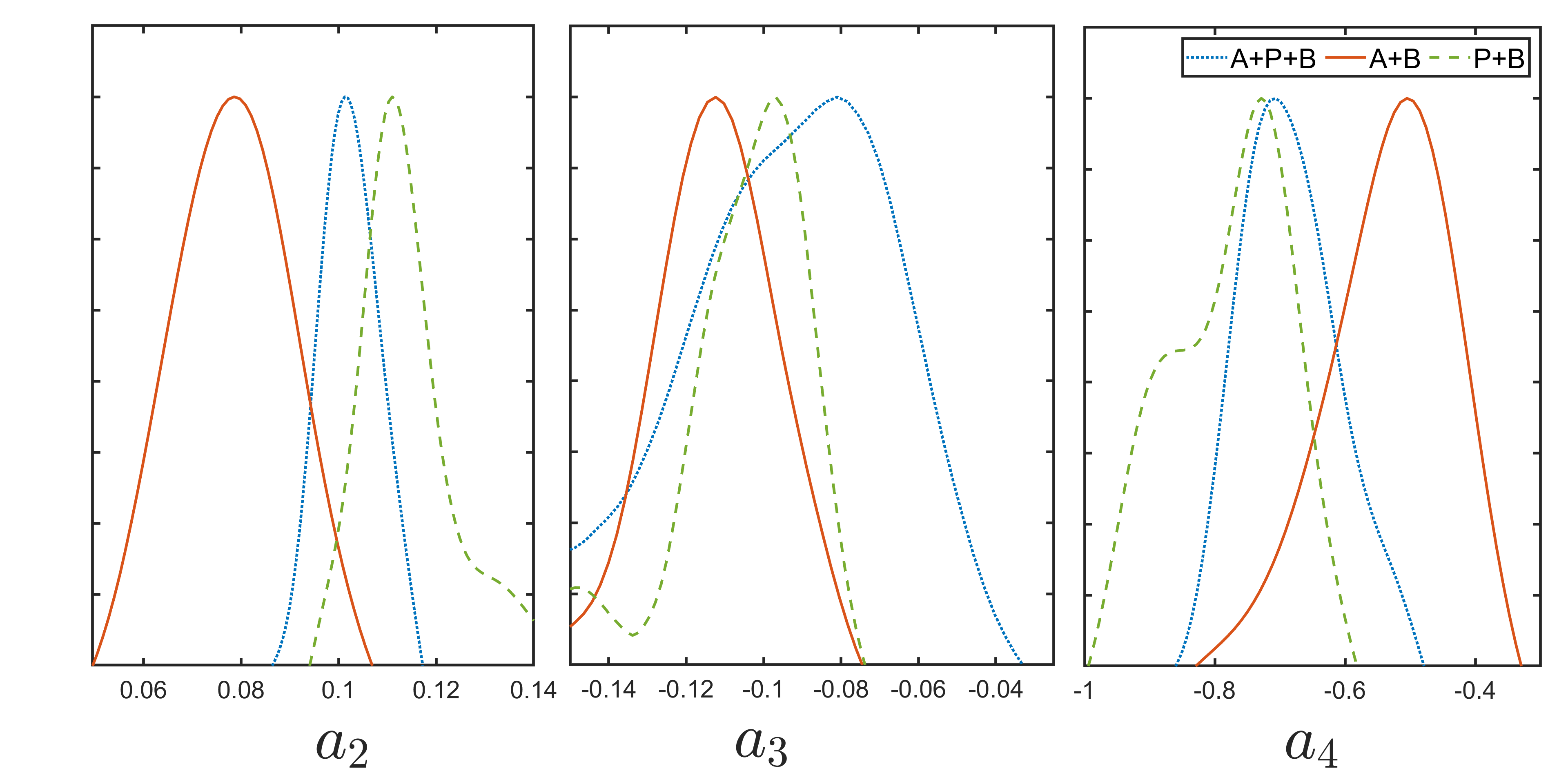}
    \caption{A comparison of the marginal likelihoods for the coefficients $\{a_2,a_3,a_4\}$ with different underlying joint $n_s-
    \alpha$ distribution. The different analyses are with ACTPol and BICEP data (red line), ACTPol,BICEP and Planck data (blue dots), and Planck with BICEP data (green dash). In these panels the distribution is shown over the entire prior, whereas in the previous ones we zoomed in on the primary peaks and the $68\%$ and $95\%$ likelihood parameter space. We can clearly see the non-linear nature of the PPS to inflationary correlation. This is evident from the density functions being Gaussian mixtures, as opposed to some single Gaussian distribution we are used to seeing.} 
    \label{fig:VisComparison}
\end{figure}

The results are conveniently summarized in table~\ref{tab:RESULTS}. As expected the uncertainties for the results of the ACTPol data alone are larger, since the data itself is less constraining. The marginal likelihoods are graphically compared in figure~\ref{fig:VisComparison}. An interesting observation is that the second coefficient $a_2$, which corresponds to the second derivative of the potential at $\phi_{\mathrm{CMB}}$ and the second slow roll parameter $\eta$ is positive in these cases. This should not be surprising as these models have enhanced scale depepndence \cite{Ben-Dayan:2009fyj}, and $n_s$ are not well approximated by usual slow-roll expression.
\begin{table}[t]
    \centering
    \begin{tabular}{||l||c | c| c ||}
    \hline
        Coefficient & ACTPol+Planck +BICEP& ACTPol+BICEP& Planck+BICEP\\
        \hline
        $a_2$ & $0.108\pm 0.011$ &$0.078\pm 0.034$&$0.111\pm 0.016 $\\
        \hline
        $a_3$&$-0.114\pm 0.014$ & $-0.112\pm 0.047$&$-0.099 \pm 0.033$\\
        \hline
        $a_4$&$-0.650 \pm 0.085$ & $-0.496\pm 0.219$&$-0.723 \pm 0.177 $ \\
        \hline
        $n_s$ (mean)&$0.973$ & $0.980$&$0.973 $ \\
        \hline
        $\alpha$ (mean)&$0$ & $0.066$&$-0.009$ \\
        \hline
    \end{tabular}
    \caption{The results for the most likely coefficients $\{a_2,a_3,a_4\}$ for a sixth degree polynomial inflationary potential. These potentials yield a tensor-to-scalar ratio of $r\simeq0.03$, while also producing the correct scalar index and index running. All analyses were combined with the latest BICEP dataset. }
    \label{tab:RESULTS}
\end{table}

\section{Discussion}
Features of the models we study include a significant running of the spectral index and a high tensor-to-scalar ratio $(r\lesssim0.03)$. Interestingly, up until now, the index running was measured very coarsely and the findings supported a vanishingly small $\alpha$. However, in the foreseeable future $\alpha$ will be measured much more precisely. For this to happen higher multipole moments should be reliably measured. With that said, already at this time, we see a possible disagreement between higher multipole analysis (ACTPol), and lower multipole analysis (Planck, WMAP etc.). This disagreement supports a finding of an appreciable running, thus makes our models more likely. Small-field models do not have good analytical predictors. Thus, we used an ANN-Encoder approach to generate pairs of coefficients-observables sets. This facilitates a rapid MCMC analysis which was previously computationally excluded for small-field models. We found that these models can support the current observational findings, while generating a GW signal with tensor-to-scalar ratio $r\sim 0.03$.  \\

The usual criticism of such a phenomenological approach cites fine-tuning. The argument is that the coefficient of these models are so fine tuned as to virtually nullify the likelihood of the models. However, we have shown that the finest tuning required for this model is at the level of $\sim 1/60$ which is no worse than most other models. \\

~As can be seen in table~\ref{tab:RESULTS}, the ACTPol+BICEP analysis disagrees with the other two analysis by several standard deviations. Moreover, considering complementary probes as mentioned in the introduction, it disagrees with suprenovae lensing available data \cite{Ben-Dayan:2013eza,Ben-Dayan:2014iya,Ben-Dayan:2015zha} and the $N_{eff}$ analysis \cite{Ben-Dayan:2019gll}, though the latter is based on some extrapolation. This may imply that the ACTPol result is due to some unknown systematic error. It would be interesting to combine these constraints into the analysis for a better determination of the most likely $n_s,\alpha$ or the potential parameters $a_2-a_4$.
We expect that future missions produce more specific data, further constraining the running of the scalar index. Thus these models are testable in the near future.

\section*{Acknowledgements}
RB is supported by the German Research Foundation through a German-Israeli Project Cooperation (DIP) grant ``Holography and the Swampland.''\\
IW thanks Roberto Trotta for useful discussions.


\begin{thebibliography}{9}
\bibitem{Penzias:1965wn}
A.~A.~Penzias and R.~W.~Wilson,
``A Measurement of excess antenna temperature at 4080-Mc/s,''
Astrophys. J. \textbf{142} (1965), 419-421


\bibitem{Starobinsky:1980te}
A.~A.~Starobinsky,
``A New Type of Isotropic Cosmological Models Without Singularity,''
Phys. Lett. B \textbf{91} (1980), 99-102

\bibitem{Sato:1980yn}
K.~Sato,
``First Order Phase Transition of a Vacuum and Expansion of the Universe,''
Mon. Not. Roy. Astron. Soc. \textbf{195} (1981), 467-479
NORDITA-80-29.

\bibitem{Guth:1980zm}
A.~H.~Guth,
``The Inflationary Universe: A Possible Solution to the Horizon and Flatness Problems,''
Phys. Rev. D \textbf{23} (1981), 347-356


\bibitem{Linde:1981mu}
A.~D.~Linde,
``A New Inflationary Universe Scenario: A Possible Solution of the Horizon, Flatness, Homogeneity, Isotropy and Primordial Monopole Problems,''
Phys. Lett. B \textbf{108} (1982), 389-393

\bibitem{Albrecht:1982wi}
A.~Albrecht and P.~J.~Steinhardt,
``Cosmology for Grand Unified Theories with Radiatively Induced Symmetry Breaking,''
Phys. Rev. Lett. \textbf{48} (1982), 1220-1223




\bibitem{Mukhanov:1981xt}
V.~F.~Mukhanov and G.~V.~Chibisov,
``Quantum Fluctuations and a Nonsingular Universe,''
JETP Lett. \textbf{33} (1981), 532-535

\bibitem{Hawking:1982cz}
S.~W.~Hawking,
``The Development of Irregularities in a Single Bubble Inflationary Universe,''
Phys. Lett. B \textbf{115} (1982), 295

\bibitem{Starobinsky:1982ee}
A.~A.~Starobinsky,
``Dynamics of Phase Transition in the New Inflationary Universe Scenario and Generation of Perturbations,''
Phys. Lett. B \textbf{117} (1982), 175-178

\bibitem{Guth:1982ec}
A.~H.~Guth and S.~Y.~Pi,
``Fluctuations in the New Inflationary Universe,''
Phys. Rev. Lett. \textbf{49} (1982), 1110-1113

\bibitem{Bardeen:1983qw}
J.~M.~Bardeen, P.~J.~Steinhardt and M.~S.~Turner,
``Spontaneous Creation of Almost Scale - Free Density Perturbations in an Inflationary Universe,''
Phys. Rev. D \textbf{28} (1983), 679



\bibitem{Bartolo:2004if}
N.~Bartolo, E.~Komatsu, S.~Matarrese and A.~Riotto,
``Non-Gaussianity from inflation: Theory and observations,''
Phys. Rept. \textbf{402} (2004), 103-266
doi:10.1016/j.physrep.2004.08.022
[arXiv:astro-ph/0406398 [astro-ph]].




\bibitem{Grishchuk:1974ny}
L.~P.~Grishchuk,
``Amplification of gravitational waves in an istropic universe,''
Zh. Eksp. Teor. Fiz. \textbf{67} (1974), 825-838

\bibitem{Starobinsky:1979ty}
A.~A.~Starobinsky,
``Spectrum of relict gravitational radiation and the early state of the universe,''
JETP Lett. \textbf{30} (1979), 682-685

\bibitem{Abbott:1984fp}
L.~F.~Abbott and M.~B.~Wise,
``Constraints on Generalized Inflationary Cosmologies,''
Nucl. Phys. B \textbf{244} (1984), 541-548



\bibitem{Bennett:1996ce}
C.~L.~Bennett, A.~Banday, K.~M.~Gorski, G.~Hinshaw, P.~Jackson, P.~Keegstra, A.~Kogut, G.~F.~Smoot, D.~T.~Wilkinson and E.~L.~Wright,
``Four year COBE DMR cosmic microwave background observations: Maps and basic results,''
Astrophys. J. Lett. \textbf{464} (1996), L1-L4
[arXiv:astro-ph/9601067 [astro-ph]].


\bibitem{Hinshaw:2012aka}
G.~Hinshaw \textit{et al.} [WMAP],
``Nine-Year Wilkinson Microwave Anisotropy Probe (WMAP) Observations: Cosmological Parameter Results,''
Astrophys. J. Suppl. \textbf{208} (2013), 19
[arXiv:1212.5226 [astro-ph.CO]].



\bibitem{Akrami:2018odb}
Y.~Akrami \textit{et al.} [Planck],
``Planck 2018 results. X. Constraints on inflation,''
Astron. Astrophys. \textbf{641} (2020), A10
[arXiv:1807.06211 [astro-ph.CO]].



\bibitem{Martin:2013tda}
J.~Martin, C.~Ringeval and V.~Vennin,
``Encyclop\ae{}dia Inflationaris,''
Phys. Dark Univ. \textbf{5-6} (2014), 75-235
[arXiv:1303.3787 [astro-ph.CO]].

\bibitem{Martin:2014lra}
J.~Martin, C.~Ringeval, R.~Trotta and V.~Vennin,
``Compatibility of Planck and BICEP2 in the Light of Inflation,''
Phys. Rev. D \textbf{90} (2014) no.6, 063501
[arXiv:1405.7272 [astro-ph.CO]].

\bibitem{Lyth:1996im}
D.~H.~Lyth,
``What would we learn by detecting a gravitational wave signal in the cosmic microwave background anisotropy?,''
Phys. Rev. Lett. \textbf{78} (1997), 1861-1863
doi:10.1103/PhysRevLett.78.1861
[arXiv:hep-ph/9606387 [hep-ph]].



\bibitem{Garg:2018reu}
S.~K.~Garg and C.~Krishnan,
JHEP \textbf{11}, 075 (2019)
doi:10.1007/JHEP11(2019)075
[arXiv:1807.05193 [hep-th]].

\bibitem{Ben-Dayan:2018mhe}
I.~Ben-Dayan,
``Draining the Swampland,''
Phys. Rev. D \textbf{99}, no.10, 101301 (2019)
[arXiv:1808.01615 [hep-th]].

\bibitem{Palti:2019pca}
E.~Palti,
``The Swampland: Introduction and Review,''
Fortsch. Phys. \textbf{67} (2019) no.6, 1900037
[arXiv:1903.06239 [hep-th]].

\bibitem{Kehagias:2019iem}
A.~Kehagias and A.~Riotto,
``A Note on the Swampland Distance Conjecture,''
Fortsch. Phys. \textbf{68} (2020) no.1, 1900099
[arXiv:1911.09050 [hep-th]].



\bibitem{Ben-Dayan:2009fyj}
I.~Ben-Dayan and R.~Brustein,
``Cosmic Microwave Background Observables of Small Field Models of Inflation,''
JCAP \textbf{09} (2010), 007
[arXiv:0907.2384 [astro-ph.CO]].


\bibitem{Hotchkiss:2011gz}
S.~Hotchkiss, A.~Mazumdar and S.~Nadathur,
``Observable gravitational waves from inflation with small field excursions,''
JCAP \textbf{02} (2012), 008
[arXiv:1110.5389 [astro-ph.CO]].



\bibitem{Wolfson:2019rwd}
I.~Wolfson and R.~Brustein,
``Small field models of inflation that predict a tensor-to-scalar ratio $r=0.03$,''
Phys. Rev. D \textbf{100} (2019) no.4, 043522
[arXiv:1903.11820 [astro-ph.CO]].


\bibitem{ACT:2020gnv}
S.~Aiola \textit{et al.} [ACT],
``The Atacama Cosmology Telescope: DR4 Maps and Cosmological Parameters,''
JCAP \textbf{12} (2020), 047
[arXiv:2007.07288 [astro-ph.CO]].


\bibitem{Wolfson:2016vyx}
I.~Wolfson and R.~Brustein,
``Small field models with gravitational wave signature supported by CMB data,''
PLoS One \textbf{13} (2018), 1-22
[arXiv:1607.03740 [astro-ph.CO]].

\bibitem{Wolfson:2018lel}
I.~Wolfson and R.~Brustein,
``Likelihood analysis of small field polynomial models of inflation yielding a high Tensor-to-Scalar ratio,''
PLoS One \textbf{14} (2019), e0215287
[arXiv:1801.07057 [astro-ph.CO]].


\bibitem{BICEP:2021xfz}
P.~A.~R.~Ade \textit{et al.} [BICEP and Keck],
``Improved Constraints on Primordial Gravitational Waves using Planck, WMAP, and BICEP/Keck Observations through the 2018 Observing Season,''
Phys. Rev. Lett. \textbf{127} (2021) no.15, 151301
[arXiv:2110.00483 [astro-ph.CO]].


\bibitem{Tristram:2021tvh}
M.~Tristram, A.~J.~Banday, K.~M.~G\'orski, R.~Keskitalo, C.~R.~Lawrence, K.~J.~Andersen, R.~B.~Barreiro, J.~Borrill, L.~P.~L.~Colombo and H.~K.~Eriksen, \textit{et al.}
``Improved limits on the tensor-to-scalar ratio using BICEP and Planck data,''
Phys. Rev. D \textbf{105} (2022) no.8, 083524
[arXiv:2112.07961 [astro-ph.CO]].

\bibitem{Sailer:2021yzm}
N.~Sailer, E.~Castorina, S.~Ferraro and M.~White,
``Cosmology at high redshift \textemdash{} a probe of fundamental physics,''
JCAP \textbf{12}, no.12, 049 (2021)
[arXiv:2106.09713 [astro-ph.CO]].

\bibitem{Chluba:2012we}
J.~Chluba, A.~L.~Erickcek and I.~Ben-Dayan,
``Probing the inflaton: Small-scale power spectrum constraints from measurements of the CMB energy spectrum,''
Astrophys. J. \textbf{758}, 76 (2012)
[arXiv:1203.2681 [astro-ph.CO]].

\bibitem{Ben-Dayan:2013eza}
I.~Ben-Dayan and T.~Kalaydzhyan,
``Constraining the primordial power spectrum from SNIa lensing dispersion,''
Phys. Rev. D \textbf{90}, no.8, 083509 (2014)
[arXiv:1309.4771 [astro-ph.CO]].

\bibitem{Ben-Dayan:2014iya}
I.~Ben-Dayan,
``Lensing dispersion of SNIa and small scales of the primordial power spectrum,''
[arXiv:1408.3004 [astro-ph.CO]].

\bibitem{Ben-Dayan:2015zha}
I.~Ben-Dayan and R.~Takahashi,
``Constraints on small-scale cosmological fluctuations from SNe lensing dispersion,''
Mon. Not. Roy. Astron. Soc. \textbf{455}, no.1, 552-562 (2016)
[arXiv:1504.07273 [astro-ph.CO]].

\bibitem{Ben-Dayan:2019gll}
I.~Ben-Dayan, B.~Keating, D.~Leon and I.~Wolfson,
``Constraints on scalar and tensor spectra from $N_{eff}$,''
JCAP \textbf{06}, 007 (2019)
[arXiv:1903.11843 [astro-ph.CO]].

\bibitem{Torrado:2020dgo}
J.~Torrado and A.~Lewis,
``Cobaya: Code for Bayesian Analysis of hierarchical physical models,''
JCAP \textbf{05} (2021), 057
[arXiv:2005.05290 [astro-ph.IM]].

\bibitem{Lewis:1999bs}
A.~Lewis, A.~Challinor and A.~Lasenby,
``Efficient computation of CMB anisotropies in closed FRW models,''
Astrophys. J. \textbf{538} (2000), 473-476
[arXiv:astro-ph/9911177 [astro-ph]].

\bibitem{ACT:2020frw}
S.~K.~Choi \textit{et al.} [ACT],
``The Atacama Cosmology Telescope: a measurement of the Cosmic Microwave Background power spectra at 98 and 150 GHz,''
JCAP \textbf{12} (2020), 045
[arXiv:2007.07289 [astro-ph.CO]].

\bibitem{Santurkar:2018}
S.~Santurkar, D.~Budden and N.~Shavit,
``Generative Compression,''
Picture Coding Symposium (PCS), (2018) 258-262.

\bibitem{Chen:2016}
X.~Chen, \textit{et al.},
``Variational Lossy Autoencoder,''
[arXiv:1611.02731]

\bibitem{Khemakhem:2019}
I.~Khemakhem,  D.~P.~Kingma,  R.~P.~Monti and A.~Hyv{\"a}rinen,
``Variational Autoencoders and Nonlinear ICA: A Unifying Framework,''
[arXiv:1907.04809]

\bibitem{kingma:2019}
D.~P.~Kingma,  and M.~Welling,
``An Introduction to Variational Autoencoders,''
[arXiv:1906.02691]


\bibitem{Wolfson:2021zsw}
I.~Wolfson,
``Analytic correlation of inflationary potential to power spectrum shape: limits of validity, and `no-go' for small field model analytics,''
JCAP \textbf{01} (2022) no.01, 036
[arXiv:2110.10557 [astro-ph.CO]].

\bibitem{Gelman:1992zz}
A.~Gelman and D.~B.~Rubin,
``Inference from Iterative Simulation Using Multiple Sequences,''
Statist. Sci. \textbf{7} (1992), 457-472


\end{thebibliography}
\end{document}